\documentclass[iop]{emulateapj}

\usepackage{natbib}
\usepackage{multirow}
\usepackage{graphicx}

\usepackage{natbib}

\slugcomment{PASP Accepted}

\shorttitle{ECR of Sirius with a CID}
\shortauthors{Batcheldor  et al.}

\begin{document}

\title{Extreme Contrast Ratio Imaging of Sirius with a Charge Injection Device}


\author{
D. Batcheldor\altaffilmark{1},
R. Foadi\altaffilmark{1},
C. Bahr\altaffilmark{1},
J. Jenne\altaffilmark{2},
Z. Ninkov\altaffilmark{3},
S. Bhaskaran\altaffilmark{2}
\&
T. Chapman\altaffilmark{2}}

\altaffiltext{1}{Department of Physics and Space Sciences, 
Florida Institute of Technology, 150 West University Boulevard, 
Melbourne, FL 32901, USA. Email: dbatcheldor@fit.edu}
\altaffiltext{2}{Thermo Scientific - CIDTEC, 101 Commerce Boulevard, Liverpool, NY 13088, USA}
\altaffiltext{3}{Center for Imaging Science, Rochester Institute of Technology, 
84 Lomb Memorial Drive, Rochester, NY 14623-5603, USA}


%\altaffiltext{1}{Visiting Astronomer, Cerro Tololo Inter-American Observatory.
%CTIO is operated by AURA, Inc.\ under contract to the National Science


\begin{abstract}
The next fundamental steps forward in understanding our place in the universe could be a result of advances in extreme contrast ratio (ECR) imaging and point spread function (PSF) suppression. For example, blinded by quasar light we have yet to fully understand the processes of galaxy and star formation and evolution, and there is an ongoing race to obtain a direct image of an exoearth lost in the glare of its host star. To fully explore the features of these systems we must perform observations in which contrast ratios of at least one billion can be regularly achieved with sub 0\farcs1 inner working angles. Here we present the details of a latest generation 32-bit charge injection device (CID) that could conceivably achieve contrast ratios on the order of one billion. We also demonstrate some of its ECR imaging abilities for astronomical imaging. At a separation of two arc minutes, we report a direct contrast ratio of $\Delta{m_v}=18.3, \log{(CR)}=7.3$, or 1 part in 20 million, from observations of the Sirius field. The atmospheric conditions present during the collection of this data prevented less modest results, and we expect to be able to achieve higher contrast ratios, with improved inner working angles, simply by operating a CID at a world-class observing site. However, CIDs do not directly provide any PSF suppression. Therefore, combining CID imaging with a simple PSF suppression technique like angular differential imaging, could provide a cheap and easy alternative to the complex ECR techniques currently being employed. 
\end{abstract}



\section{Introduction}

There is an increasing demand for instrumentation and techniques capable of providing contrast ratios (CRs) not directly achievable by metal-oxide-semiconductor (MOS) arrays equipped with standard 16-bit analog-to-digital converters, i.e., $\log{(CR)}\gtrsim5~(\Delta{m}\sim12.5)$. This demand spans almost the entire astronomical mass spectrum, from the direct imaging of exoearths, to the host galaxies of quasars. The case of the Sun-Earth system highlights the extreme contrast ratio (ECR) problem to be overcome when attempting to image an exoearth. The Sun-Earth apparent magnitude difference is 22.9, which corresponds to an ECR of $\log{(CR)}={9.2}$. The case of the $\sim4M_\oplus$ $\tau$-Ceti e  \citep{2013A&A...551A..79T} gives similar results, where the magnitude difference between this planet and $\tau$-Ceti itself would be 21.5 magnitudes (assuming a 0\farcs16 semi-major axis, 50\% phase and an Earth-like albedo). This corresponds to an ECR of $\log{(CR)}={8.6}$. The case of quasar hosts is less well understood because of the challenges associated with getting deep images of the underlying galaxy. However, consider the case of 3C273, a bright ($m_v = 12.9$, $M_v = -26.7$) quasar hosted by a giant galaxy \citep{2003AJ....125.2964M}. The apparent isophotes of the $M_v$ = -23.8 host galaxy extend to a radius of $\sim$25 kpc and $m_v\sim24$ . Consequently, it is clear that not all of the 3C273 host galaxy has been detected because the isophotes of less massive early-type galaxies like M87 extend out past 60 kpc with $m_v > 25$ \citep{2005AJ....129.2628L}. It is likely that, in the past, M87 hosted a quasar as powerful as 3C273. Therefore, the comprehensive study of highly accreting black holes at the centers of large host galaxies requires ECRs of at least $\log{(CR)}={8}$ if the large scale morphology of host galaxies is to be determined. 

The direct imaging of an Earth-like planet around another star is probably the most compelling aim of contemporary astrophysics with a rich discovery space, and a whole host of open questions. However, the questions about quasar host galaxy morphologies highlight the uncertain nature of the relationship between nuclear activity and the evolution of galaxies in general. The maturation of ECR astronomy will therefore allow us to better understand the connections between supermassive black holes and their host galaxies \citep{1995ARA&A..33..581K,1998AJ....115.2285M,2000ApJ...539L...9F,2000ApJ...539L..13G,2001ApJ...563L..11G,2003ApJ...589L..21M} and the observational biases \citep{2010ApJ...711L.108B,2011ApJ...734...92J},  and determine whether supermassive black holes regulate galaxy evolution or vice versa. Previous active galactic nuclei (AGN) host studies have dealt with bright central sources by making high contrast ratio observations in the near infrared \citep{2008A&A...478..311S} and in Type 2 (obscured) AGN \citep{2003MNRAS.346.1055K,2008ApJ...675.1025S}. In addition, the saturated AGN point-spread function (PSF) has been modeled from short unsaturated exposures \citep{2003MNRAS.340.1095D,2004MNRAS.355..196F} and the bright central source has been occulted using a coronagraph \citep[e.g.,][]{2003AJ....125.2964M}. While these studies have had some success in determining the properties of host galaxies, they use techniques that are still limited to $\log{(CR)}\sim{6}$. 

Exoearths and quasar hosts aside, there are many other areas of compelling astrophysics, mostly involving the detection of faint extended emission or nearby faint companions around otherwise bright stars, that will benefit from ECR imaging. For example, what is the binarity fraction of massive stars and the incidence of sub-stellar companions \citep[e.g.][]{2005AJ....130.1845L}? Does the apparent universality of the companion mass-ratio distribution continue below binary mass ratios of 0.02 \citep[e.g.,][]{2013A&A...553A.124R}? Could faint stars in large mass-ratio binaries with small spatial separations fall below the detection limits set by contrast ratios and inner working angles, or is there a mass-ratio limit based on the way multiple star systems form \citep[e.g.][]{2010MNRAS.405.2439O}? In addition, if there is a significant population of faint cooler low mass stars either in binary systems, or within clusters populated by otherwise massive stars, how would this effect the universality of the initial mass function \citep[e.g.,][]{2001MNRAS.322..231K,2002ApJ...579..275S}? A simple ECR survey of stellar systems could make significant progress on these questions, and potentially (as a by-product) find planetary systems, debris disks, faint nebulosity from pre- or post-main sequence evolution, and aid in the characterization of protostellar and secondary debris disks and exozodical clouds \citep{1999ApJ...525L..53W,2006ApJ...650..414S,2008ApJ...686..637S}. 

Separating the signals of relatively bright and faint targets is a significant challenge for standard MOS arrays because, in addition to charge bleeding and the PSF suppression required, the directly achievable CRs are primarily determined by the full well depth of the pixels and limited to $\log{(CR)} <5$. Therefore, a host of high contrast ($5 < \log{(CR)} <7$) ratio imaging and PSF suppression techniques have been developed. Scientific CMOS detectors developed by Fairchild and incorporated into Andor, PCO, and Hamamatsu imaging devices, use low-read noise and high frame rates to reach $\log{(CR)\sim5}$ \citep{2004SPIE.5274..194K}. Teledyne's H$x$RG family of sensors may also be able to deliver competitive contrast ratios \citep[e.g.,][]{2011ASPC..437..383B}. Coronagraphy \citep[e.g.,][]{2001AJ....121..525S}, nulling interferometry \citep{2003SPIE.4852..443L}, integral field spectral deconvolution \citep{2002ApJ...578..543S}, and spectral differential imaging \citep{2012PASP..124..454I} are all techniques that are primarily concerned with PSF suppression. Each has their own advantages and disadvantages. For example, the need for high wave-front quality, precise pointing control, additional support structures, and multiple apertures, all lead to significant costs and complications. Software based PSF modeling like Locally Optimized Combination of Images \citep{2007ApJ...660..770L} provide some incremental gains in faint point source detection when used with other methods, and the Karhunen-Loeve Image Projection approach used by the Archival Legacy Investigation of Circumstellar Environments project \citep{2015AAS...22534921S} are still reliant on coronagraphic observations. Ground-based angular differential imaging \citep{2006ApJ...641..556M} and space-based roll subtraction \citep{2005AJ....130.1845L,2010hstc.workE..15S} are two less complicated techniques that can help mitigate the effect of the PSF and enable high contrast ratio imaging as long as a suitably stable PSF can be achieved. 

In this paper we present preliminary contrast ratio results from the on-sky testing of a latest generation charge-injection device (CID); the SpectraCAM XDR (SXDR) supplied by ThermoFisher Scientific Inc. This 32-bit device has the potential to achieve direct contrast ratios of up to $\log{(CR)}={9}$ but does not address PSF suppression. While we do not present a comprehensive set of performance measures here, CIDs could greatly improve ECR astronomy in general; in the future CIDs could provide a wealth of valuable data on a whole host of important astronomical objects. In \S~\ref{cids} we provide an introduction to CIDs. In \S~\ref{sxdr} we introduce the SXDR. In \S~\ref{obs} we present the results of on-sky testing of the SXDR. \S~\ref{dis} discusses these results and the future of CIDs for ECR astronomy. \S~\ref{cons} concludes. 

\begin{figure}
\plotone{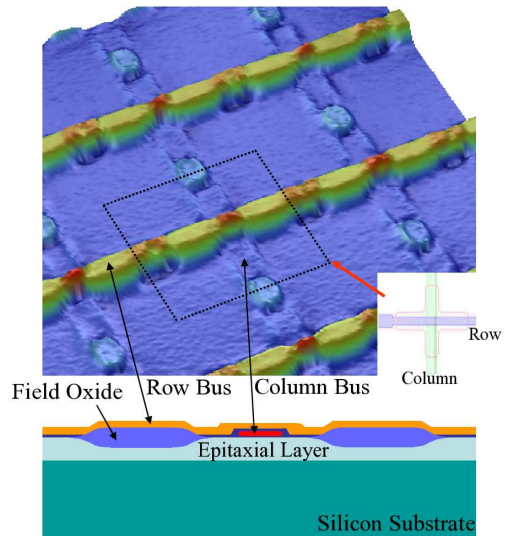} 
\caption{The simple crossed-cell CID pixel. [top] Interferogram showing the surface row bus (highest features in yellow, orange, and red). [bottom] Cross section showing the row bus in orange overlaid on the column bus in red. Adapted from \cite{2008SPIE.7055E..19B}. This design is used in the majority of passive pixel CID detectors.
}
\label{fig:pixel}
\end{figure}

\section{Charge Injection Devices}\label{cids}

The CID is an array of X-Y addressable photosensitive MOS capacitor elements invented at General Electric in 1974 \citep{g1974method}. A pixel consists of two MOS capacitors formed from continuous polysilicon strips laid along the columns and rows (Figure~\ref{fig:pixel}). Field oxide ensures pixel separation, and the continuous bus architecture allows individual pixels control from the edge of the array. In each pixel one electrode is attached to the column bus and one to the row bus. This allows each bus to act as either a sense or a storage node. The process of sensing charge on a pixel is accomplished by transferring signal from the column to the row (from the storage to the sense node). 

\begin{figure}
\plotone{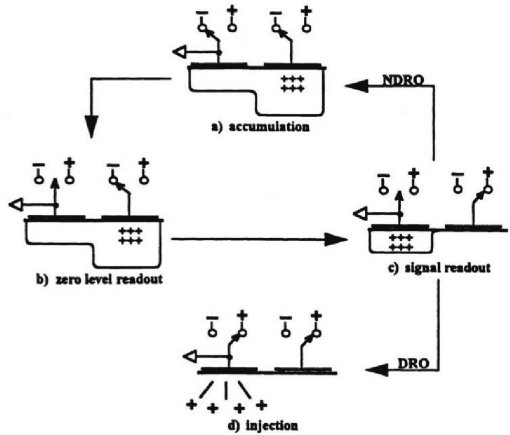}
\caption{Four-stage operation of an early CID from \cite{1994SPIE.2198..868N}. The read-out scheme follows signal accumulation, zero level read-out, then either a non-destructive read-out or destructive read-out, and charge injection. A post-injection pixel returns to the accumulation mode. In each of the four panels the sense node is on the left and the storage node on the right. 
}
\label{fig:readout}
\end{figure}

The basic operation of a CID pixel is shown in Figure~\ref{fig:readout} \citep{1994SPIE.2198..868N}. To accumulate the holes generated by incident photons in a pixel (Figure~\ref{fig:readout} stage [a]; accumulation) both the sense and storage nodes are connected to a negative potential. After the predetermined time, the sense node is then floated while the zero-level is measured (Figure~\ref{fig:readout}  stage [b]; zero level read). The charge under the storage node is then pushed to the sense node by connecting the storage node to a positive potential. The generated charge, now under the sense node, is then measured and compared to the zero level readout (Figure~\ref{fig:readout} stage [c]; signal readout). The number of incident photons is proportional to the difference between the two measured values. From here, two options are available; either the sensed charge is driven back to the storage node by re-creating the negative potential at that node, i.e., enabling a non-destructive read out (NDRO) before continued accumulation, or both nodes are set to a positive potential and the holes are ÒinjectedÓ back into the substrate (Figure~\ref{fig:readout} stage [d]; injection). After charge injection occurs a pixel returns to its accumulation mode. Signal readout on any pixel does not affect the state of other pixels in the array, and the readout does not require injection of the signal charge or transfer of charge out of the pixel site. 

Due to the field oxide, the readout architecture, and the nature of the underlying charge collection architecture, CIDs are inherently anti-blooming. CIDs are also well suited to space; they are remarkably resistant to gamma rays and neutrons \citep[e.g.,][]{2001RScI...72..713M} and have radiation tolerance past 5 Mrad \citep{2008SPIE.7055E..19B}. In addition, CIDs use NDROs that minimize the effects of cosmic rays because count rates are monitored and radiation events are removed from the final photon flux. These features are also present in CMOS and HxRG devices. 

To date, CCDs have been preferred to CIDs (and standard CMOS) due to comparative read noise levels. CCDs typically have a read noise of $<10e^-$, while early CIDs had $\sim600e^-$ \citep{1981ApOpt..20.3189M,1986SPIE..686...66M}. This was partly a result of vertical shift register controlled multiplexers at the end of each row. \cite{michon1987cid} described replacing the multiplexers with pre-amplifier per row (PPR) architectures, and \cite{doi:10.1117/12.148589} suggested narrowing the polysilicon crossovers to reduce pattern noise. These measures were introduced to the CID-38 where read noise was reduced to $\sim250e^-$ \citep{1994SPIE.2172..180N}. NDROs provide a $\sqrt{N}$ improvement in read noise (where $N$ is the number of reads) and the CID-38 managed to achieve a read noise of ~20e- with N = 100 \citep{1994SPIE.2198..868N}. \cite{1995SPIE.2415..292E} presented the pre-amplifier per pixel (PPP) architecture, and \citet{1995SPIE.2518..397K} suggested replacing the shift registers with random access decoders. These were incorporated into the CID-84, CID-85 and CID-86 \citep{1996SPIE.2654...29Z} by CID Technologies Inc., a subsidiary of what is now Thermo-Fisher Scientific. However, these devices still experienced significant clocking overheads when accessing random pixels. The CID-810 incorporated high readout speed to reduce these clocking overheads \citep{1998SPIE.3356.1036M,1999ASPC..164..392K} and \cite{2004SPIE.5167...94M} introduced the PPP based CID-817 with a read noise of 50$e^-$. Despite these advances it has been more than 20 years since a CID was directly used for astronomical observations \citep{1994SPIE.2198..868N}. The latest generation of commercially available CIDs is the SXDR. This device has comparable read noise levels to CCDs when used with multiple NDROs. 

\begin{figure}
\plotone{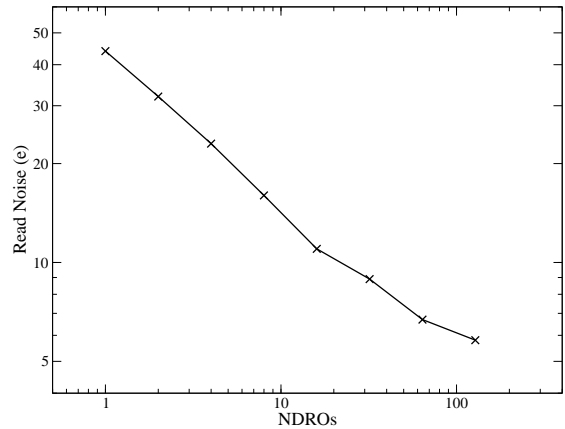}
\caption{SXDR read noise vs. number of NDROs (N). The lowest read noise is 5.8 $e^-$ RMS for N = 128.
}
\label{fig:noise}
\end{figure}

\section{The SpectraCAM XDR}\label{sxdr}

The SXDR used in this study incorporates a 2048 x 2048 12\micron~pixel Random Access (RACID) CID820 sensor with PPP. The pixels have linear (within 2\%) full wells of 268k $e^-$ that saturate at 305k $e^-$. The peak quantum efficiency of this front illuminated device is 48\% at 525 nm (approximate V-band). The hermetically sealed detector is cooled to -45.6$^o$ C using an ethylene glycol recirculation system. At this temperature the dark current is 5 $e^-$/s. The mean variance gives a conversion gain of 6.2 $e^-$ per ADU. In a single read the noise is 44 $e^-$ RMS. This drops to 5.8 $e^-$ RMS with 128 NDROs when the pixels are sampled at 2.1 MHz (Figure~\ref{fig:noise}). 

The SXDR uses a random access integrator (RAI) to control the data acquisition in an intensely illuminated region of interest (ROI). The ROI is simply determined by identifying the brightest pixels in a short (0.1 second) pre-exposure. The pixels in the ROI control area integrate until an NDRO signal threshold is tripped at 75\% of the full well. At that point the ROI is read out and recorded in the RAI with a time stamp. The ROI pixels then enter the charge injection phase before returning to the accumulation mode. When the total full frame exposure time is reached, the whole frame is read out. The RAI data is then merged into the full frame to create a composite image with a theoretical contrast ratio of up to $2^{32}$, $\log{(CR)}\sim 9.6$, or $\Delta{m\sim24}$. 

Typically, achievable contrast ratios ($CR_s$) are simply described in terms of $F$, the full well capacity in electrons, $s$, the detection limit factor (e.g., $2\sigma$ or $3\sigma$), and $N_r$ the read noise in electrons (Equation~\ref{equ:simple}). 

\begin{equation}\label{equ:simple}
CR_s=\frac{F}{sN_r}
\end{equation}

In the case of the CID the contrast ratio depends on the frequency at which the ROI can be read out ($t_{\rm min}$) and the total exposure time 
($t_{\rm max}$). In this case the extreme contrast ratio ($CR_e$) is given by Equation~\ref{equ:extreme}.

\begin{equation}\label{equ:extreme}
CR_e=\frac{Ft_{\rm max}}{sN_rt_{\rm min}}
\end{equation}

With the 5.8 e$^-$ read noise associated with 128 NDROs on the non-ROI pixels, with the ROI pixels being read every 0.01 seconds, and using the entire full well-capacity (305k e$^-$), a $2^{32}$-bit limited contrast ratio of $\log{(CR)}\sim 9.6$, with a $3\sigma$ detection on the faint source, could be reached in an exposure time of 11.4 seconds. This increases to 13 seconds if the bright source is limited to the linear full well of 268k e$^-$. However, in both cases an appropriate target field, with those levels of intrinsic contrast ratio, are needed.

An appropriate field on which to test these types of contrast ratios can be made in the lab environment using a laser. Previous studies by, for example, \cite{Welsch:2006ef} have shown $\log{(CR)}\sim 7$ has been achieved using a CID84 illuminated by a pulsed Piquant PDL 800-B laser. The contrast ratio is generated from the center of the laser pulse and the lens flares or internal reflections. However, much more extreme contrast ratios are naturally generated in astrophysical objects, and therefore the night sky potentially offers a much better sample of test fields. 

\begin{deluxetable}{lcccc}
\tablecaption{Details of Bright Objects in the Sirius Field\label{tab:objects}}
\tablewidth{0pt}
\tablehead{
\colhead{Name} &\colhead{Type} &\colhead{$m_v$} & \colhead{RA} & \colhead{DEC}
}
\startdata
$[$a] V* HM CMa & Flare Star & 8.9* & 06 45 19 & -16 48 09 \\   
$[$b] BD-16 1589 & K5 D & 8.5 & 06 44 59 & -16 47 51  \\ 
$[$c] \hspace{0.2cm}2MASS & \nodata & 9.1 & 06 44 43 & -16 45 40 \\ 
\hspace{0.45cm}J06444312\\
\hspace{0.55cm}-1645393\\
$[$d] Not catalogued & \nodata & 11.0$\dagger$ & 06 45 03 & -16 46 19 \\
\enddata
\tablecomments{Details of objects marked in Figure~\ref{fig:big} from SIMBAD\footnote{http://simbad.u-strasbg.fr/simbad/}. * The observed magnitude of [a] is $m_v\approx11.4.$ $\dagger$ This object is not in SIMBAD, but the estimated V-band magnitude based on these observations is $m_v\approx11.0.$}
\end{deluxetable}

The SXDR was installed on the Florida Tech 0.8 m Ortega telescope; a DFM Engineering f/8 RC Cassegrain located in Melbourne, Florida ($\lambda~28^o 03' 44.7"$N, $\phi~80^o 37' 26.9"$W). The detector has a 13 arcminute field-of-view with an image scale of 0\farcs4/pixel. However, the Florida environment is not conducive to astronomy, and the limitations imposed by the sky conditions at this observing site make achieving the potential contrast ratio a significant challenge. The sky background is approximately 19 magnitudes per square arc second \citep{2001MNRAS.328..689C}. Nevertheless, the background fields of bright stars offer the opportunity to approach the maximum direct imaging CRs that the system is capable of providing. Therefore, the test target for these preliminary observations was simply chosen to be Sirius, the background field of which is known to contain point sources fainter than 16th magnitude in the V-band \citep{1991A&amp;A...252..193B,2000A&amp;A...360..991B}.

\begin{figure}
\plotone{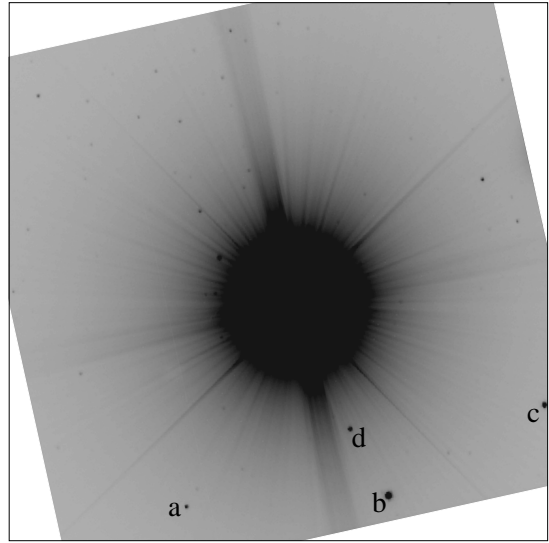}
\caption{SXDR imaged field of Sirius. North is up and East is to the left. The image is 13 arcminutes on each side. This 20 second exposure is unfiltered (approximate V-band). The signal from Sirius has not saturated the camera. The letters [a - d] identify objects that are discussed in the text and presented in Table~\ref{tab:objects}. 
}
\label{fig:big}
\end{figure}

The Sirius field was imaged using a 20 second exposure with the SXDR on March 17th, 2015, at an airmass of 1.7. The seeing was 3\farcs0 FWHM. Due to the relatively low QE of the camera we chose to maximize the system throughput by collecting the data unfiltered. However, the peak response of the SXDR is approximately equivalent to the V-band. The 13 arcminute field-of-view around Sirius is presented in Figure~\ref{fig:big}. The detector is angled at 12.4 degrees due to the necessary design of the imaging port mounting plate, which had to take into account the locations of the glycol recirculation cooling pipes. However, the field has been rotated so that North is up and East is to the left. With an ROI exposure time of 0.01 seconds, the maximum expected contrast ratio for these observations, from Equation~\ref{equ:extreme} is $\log{(CR)}\sim 7.5$ for a $3\sigma$ faint source detection.

\begin{figure*}
\plotone{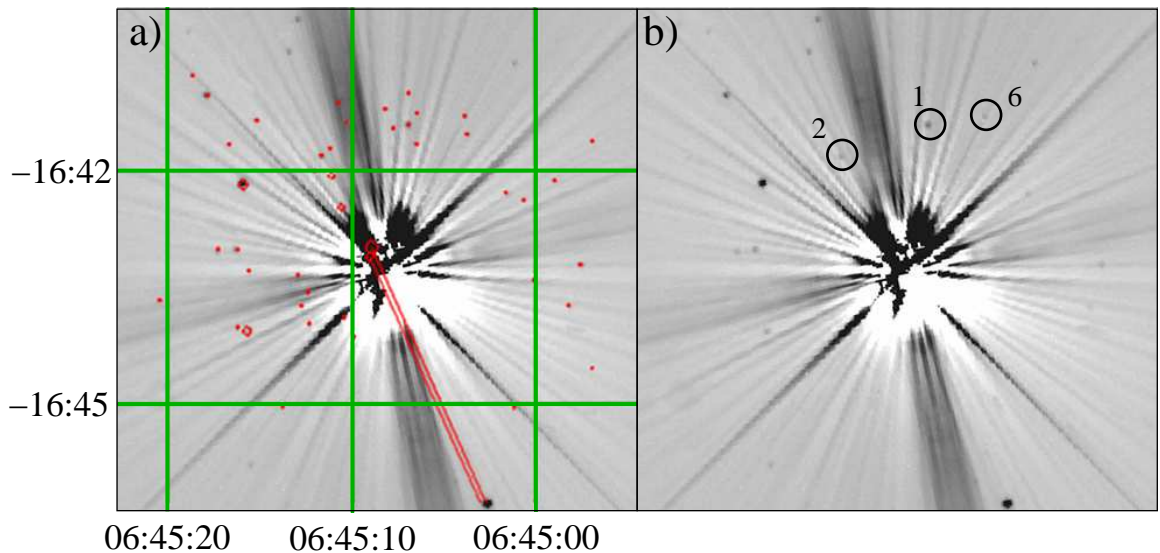}
\caption{The 3\farcm25 radius field around Sirius. [a] The SIMBAD catalog data overlaid on the SXDR image that has had a simple diffuse halo model subtracted. The red dots indicate catalogued stars, 9 of which have been duplicated due to 14 years of proper motion for Sirius (see text). [b] Identification of 3 detected stars that have had V-band magnitudes previous published by \cite{1991A&amp;A...252..193B}. 1 = [BG91] 1, $m_v=14.2$. 2 = [BG91] 2, $m_v=14.5$. 6 = [BG91] 6, $m_v=16.8$. 
}
\label{fig:compare}
\end{figure*}

\section{Results}\label{obs}

Figures~\ref{fig:big} and \ref{fig:compare} present the results from the SXDR imaging of Sirius. The signal from Sirius is not saturated. The pixels illuminated by Sirius were identified in a 0.1 second pre-exposure and assigned to an ROI. These pixels were then read out and charge injected each time they reached 75\% of their full-well. The pixels from the rest of the array were read out at the end of the exposure. The SIMBAD details of the major objects in the field (aside from Sirius) are marked [a-c] in Figure~\ref{fig:big}, and are given in Table~\ref{tab:objects}. The object marked [a] is a variable flare star, and the object marked [d] is not listed in any known catalog. Its observed position is given in Table~\ref{tab:objects} to a precision of 0\farcs5. Objects [b] and [c], as well as [BG91] 1, 2 and 6 (which are detected inside the Sirius halo, see Figure~\ref{fig:compare}), were used for an approximate ($\pm 0.5$ magnitudes) photometric calibration of the field. At the time of observation, the flare star has an approximate V-band magnitude of $m_v\approx11.4$, and object [d] $m_v\approx11.0$. 

Figure~\ref{fig:big} demonstrates two issues. First, the seeing conditions from this site cause significant scattering of the Sirius signal so that the sky background is dominated by the atmospheric halo generated. Second, the contrast in the image is not able to show the full dynamic range of the data. 
Typically, lower dynamic range 16-bit CCD data can be adequately mapped to an 8-bit display using cut-levels and logarithmic stretches. However, the SXDR data is 32-bit and presents a significant display challenge. The issue of displaying 32-bit data can be somewhat mitigated by modeling and subtracting the atmospheric halo.  

Figure~\ref{fig:compare} shows the central 6\farcm5 square field around Sirius with a simple smooth diffuse halo model subtracted from the data. In this case, a surface brightness profile was fitted to Sirius using the IRAF package {\it ellipse}. The results of this isophotal analysis were then built into a 2D halo model image using the IRAF package {\it bmodel}. The residuals between this model and the original data are what is shown in Figure~\ref{fig:compare}. Overlaid, in the left hand panel of Figure~\ref{fig:compare}, are the SIMBAD catalogued stars (red dots). The green lines show the astrometry as determined from a six star alignment with [BLC2000] (11,12,13,17,21,22). Note that [BLC2000] 11 and 13 are also 2MASS J06451562-1644043 and 2MASS J06451589-1642099, respectively. We have assumed each of these stars has had negligible proper motion over a baseline of 15 years. Our estimated astrometric uncertainty is 3", and the estimated current position of Sirius is $06^h45^m08^s, -16^o43'19"$. This is a 20" offset from the SIMBAD listed position, and consistent with 15 years of proper motion. The solid red lines show the linear proper motion vector of Sirius A and B. The slight offset from this vector is a result of the orbital motion of Sirius A about the center of mass of the binary system. 

SIMBAD reports 43 stars within a radius of 3\farcm25 around Sirius. However, 9 of these ([BG91] 1-9,  \citealt{1991A&amp;A...252..193B}) are duplicated with [BCL2000] 1-9 \citep{2000A&amp;A...360..991B} due to the different spatial positions reported by each author. The [BG91] designations use positions as offset from Sirius, whereas the positions reported for the [BCL2000] designations are absolute. There is a 15-year gap between these observations, and so the proper motion of Sirius has rendered the spatial position of the [BG91] sample obsolete. However, the [BG91] sample have estimated V-band magnitudes, whereas the [BCL2000] sample have Gunn-z magnitudes (830-950 nm sensitivity). As the V-band most closely matches the peak wavelength sensitivity of the SXDR, we adopt the [BG91] magnitudes, but use the [BCL2000] spatial positions. However, we note that the differences in the response functions between the [BG91] system and the SXDR system limit the precision with which we can perform photometry.

The three stars highlighted in Figure~\ref{fig:compare} ([BCL2000] 1, 2 and 6) are clearly identified in the field. Both [BCL2000] 1 ($m_v=14.2$) and 2 ($m_v=14.5$) are detected at $4\sigma$ above the background level, and [BCL2000] 6 ($m_v=16.8$) is detected at $3\sigma$ above the background level. When compared to the V-band magnitude of Sirius (-1.46), this corresponds to direct raw contrast ratios of $\log{(CR)}=$6.3, 6.4, and 7.3 respectively. 

\begin{figure*}
\plotone{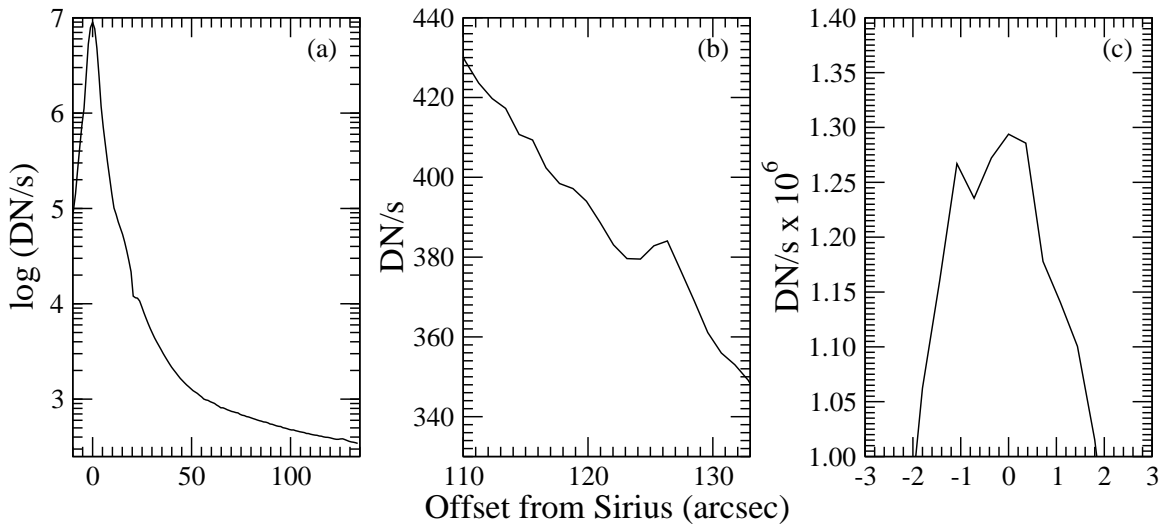}
\caption{Radial profiles of Sirius along the direction to [BCL2000] 6. The profiles in (a) and (b) are from data summed in 3x3 bins. (a) The radial profile North-West of Sirius in log data numbers per second. (b) The second of the radial profile that includes the signal from [BCL2000] 6 in linear data numbers per second. (c) The peak of the profile for Sirius itself in linear data numbers per second from un-binned data. 
}
\label{fig:profiles}
\end{figure*}

\begin{deluxetable}{lcccc}
\tablecaption{Other Known Targets in the Sirius Field\label{tab:targets}}
\tablewidth{0pt}
\tablehead{
\colhead{Name} &\colhead{Seperation (")} & \colhead{$m_v$} & \colhead{Detected} & \colhead{$\log{(CR)}$}
}
\startdata
Sirius B				      & 8\farcs1     & 8.4    & No & 3.9 \\
BD-16 1591D 			      & 39\farcs3   & 14.0 & No & 6.2 \\
$\left[{\rm BCL2000}\right] 1$ & 98\farcs7   & 14.2 & Yes & 6.3 \\
$\left[{\rm BCL2000}\right] 2$ & 80\farcs1   & 14.5 & Yes & 6.4 \\
$\left[{\rm BCL2000}\right] 3$ & 60\farcs7   & 16.4 & No & 7.1 \\
$\left[{\rm BCL2000}\right] 4$ & 71\farcs0   & 17.2 & Maybe & 7.5 \\
$\left[{\rm BCL2000}\right] 5$ & 114\farcs4 & 16.6 & Maybe & 7.2 \\
$\left[{\rm BCL2000}\right] 6$ & 125\farcs9 & 16.8 & Yes & 7.3 \\
$\left[{\rm BCL2000}\right] 7$ & 58\farcs9   & 17.6 & No & 7.6 \\
$\left[{\rm BCL2000}\right] 8$ & 107\farcs1 & 17.9 & No & 7.8 \\
$\left[{\rm BCL2000}\right] 9$ & 122\farcs6 & 17.8 & No & 7.7 \\
\enddata
\tablecomments{Targets in the Sirius field that have published V-band photometry.}
\end{deluxetable}

To demonstrate the detection of [BCL2000] 6 further, Figure~\ref{fig:profiles} shows the radial profile of Sirius along a position angle of $32^o$ west of north. This takes the profile through the position of Sirius and [BCL2000] 6. The left hand panel (Fig~\ref{fig:profiles}[a]) shows the unsaturated signal of Sirius in log data numbers per second, and demonstrates the size of the diffuse halo. [BCL2000] 6 is a barely noticeable bump at a separation of 126". The center panel (Fig~\ref{fig:profiles}[b]) highlights the region along the profile in which [BCL2000] 6 resides, where the signal can be seen more clearly. In both panel (a) and (b) the profiles have been extracted from data that has been binned 3x3. 

The right hand panel of Figure~\ref{fig:profiles} shows the peak un-binned signal from Sirius passing through the brightest pixel. Due to the nature of the read out algorithm in the ROI, this profile warrants some explanation. It has neither the flat topped form of a saturated signal, nor the gaussian profile expected from the 3" FWHM seeing conditions. The 0.1 second pre-exposure taken in order to establish which pixels will be assigned to the ROI samples the field on the order of the atmospheric scintillation timescale. Therefore, the spatial position of the brightest pixels will change within the ROI from read to read leading to the random pattern of signal observed within the ROI itself. 

\section{Discussion}\label{dis}

The successful achievement of a raw contrast ratio of $\Delta{m_v}=18.3, \log{(CR)}=7.3$, or 1 part in 20 million, is a moderate but encouraging result. It provides motivation for the continued improvement of CIDs for extreme contrast ratio astronomy applications.  We expect that these preliminary observations, at a non-ideal site using a prototype device, will be improved upon in the near future. The QE of the SXDR currently tested here, which includes the CID820 chip, is relatively low (48\% at 550nm) due to it being front illuminated. However, Thermo-Fisher Scientific is in the process of incorporating a back thinned CID821 into an SXDR type package with an estimated QE of 85\% at 550nm. This new device will incorporate a multi-stage thermo-electric cooler that will take the sensor to 100 K below ambient. This will reduce the dark current well below 5 $e^-$/s. The increased sensitivity of this next generation CID will enable these detectors to be used for more ambitious observing programs at world-class facilities with sub-arcsecond seeing. Consequently, improvements on the issues caused by the Florida atmosphere will be made by operating a back illuminated CID 821 on the 1.0-m Jacobus Kapteyn Telescope, located on La Palma, in the summer of 2016.

Another motivating factor for installing a CID at a world-class observatory is in improving the IWA simply by having better seeing conditions. Sirius B, despite being $8^{th}$ magnitude and separated by 8\farcs1, is not detected in this data. The projected separations from Sirius of [BCL2000] (2,1,6) are 64\farcs4, 87\farcs3, and 114\farcs3, respectively. These IWAs are far from the levels of performance necessary for most ECR imaging applications, and they highlight the fact that, despite good raw contrast ratio performance, CIDs do nothing for suppressing the effects of the PSF. Therefore, a combination of CIDs with easy to perform PSF subtraction methods, like angular differential imaging \citep{2006ApJ...641..556M}, would also likely make progress towards the types of ECRs being demanded by exoearth and quasar host imaging. 

With better atmospheric conditions the Sirius field offers the opportunity to further test the achievable contrast ratios. Table~\ref{tab:targets} lists all the possible targets in the field with reported V-band magnitudes, including the 3 that we detected at better than $3\sigma$ above background. In addition, there are two possible detections listed as ``maybes". In these cases a small excess above the background was detected at spatial positions coincident with the catalogue positions. However, the signal-to-noise ratio is too small for us to be confident that these excesses are not simply residual features in the PSF.

We note the serendipity in this data that is caused by [BCL2000] 6 being separated just far enough to be detected, with confidence, above the scattering diffuse halo around Sirius. Were [BCL2000] 6 10" closer, it is unlikely that it would have been detected. Conversely, at a better observing site, we easily expect to be able to detect [BCL2000] 8, which would demonstrate an observed raw contrast ratio of over 60 million ($\log{(CR)}=7.8, \Delta{m_v}=19.4$).

The imaging of the Sirius field provided here does demonstrate that the objects in this region of the celestial sphere remain largely under-studied. Considering the brightness of Sirius, this is of no surprise. While star [d] in Figure~\ref{fig:big} does not appear in any catalogue, the object itself is unlikely to be a rapid transient because it is clearly visible in previous images taken by, for example, amateur astronomer Yuuji Kitahara using a 16 inch reflector. This raises an interesting question about any bright star field. What other unknown objects are going to be in those fields? A simple ECR imaging survey of bright star fields might reveal some interesting targets. 

The data gathered by this program highlights that the field of extreme contrast ratio imaging would benefit from some development of appropriate display tools, in addition to a previously known ``knockdown" issue. While a pixel in the ROI is being reset it is not detecting photons. Therefore, some flux will be missing when the RAI is merged at the end of the exposure. When the achieved contrast ratio, based on the published relative magnitudes (20 million), is compared to the measured count rate ratio between Sirius and [BCL2000] 6 (9 million), we find the observed flux ratio is 45\% lower than expected. Such a loss, based on 2000 0.01 second samples within the longer 20 second exposure, could be accounted for with a reset and injection (knockdown) period of 300 $\mu$s. 

However, knockdown does provide some benefits for increasing the absolute raw contrast ratio range possible, as up to 45\% of the bright source flux is being inadvertently suppressed. Indeed, a future development of CID technology will include the ability to simply ``turn-off" the ROI pixels by floating the gate voltages. In this mode, a CID would essentially be able to operate as a dynamic coronagraph because a pseudo-coronagraphic ``hole" could be created anywhere on the detector. This would relax the precision acquisition pointing requirements of the telescope. However, as well as appropriate PSF mitigation techniques, precise tracking would still be required in order to ensure the flux from the bright target lands on as few pixels as possible.

In the cases where absolute photometry is needed, knockdown is something that can be calibrated and corrected for using simple standard star observations and a knockdown coefficient. The pre-exposure could also be used to determine the expected count rates independent of the knockdown, and therefore may be useful in determining the knockdown coefficient. While in most cases the flux from the bright source will be of no interest, there may be some scenarios in which those data are valuable. One example may be in the photometric monitoring of reflected light from an exoplanet. If that planet has an anisotropic albedo, that signal will need to be separated from any potential photometric variations in the host star itself, especially for small IWAs where the light travel time from the host star to the planet will be on the order of minutes. However, it is likely that differential photometry could disentangle these signals in a relative sense without the need for a precise photometric calibration. 

As well as continuing to improve the levels of read noise, a valuable feature that can be added to future generations of CIDs is the ability to assign multiple ROIs within a single frame. When considering the potential populations of faint cooler low mass stars, star clusters would make an excellent choice for further study. The massive stars in the cluster would normally limit the depth of any exposures, but a multi-ROI CID could overcome this issue and potentially reach a limiting magnitude unachievable by any other high-contrast ratio imaging technique. New limits to the initial mass function could then be placed at the low mass end, rather than extrapolated from the high mass data.  

\section{Conclusions}\label{cons}

A latest generation CID has been used to easily demonstrate a direct contrast ratio of $\Delta{m_v}=18.3, \log{(CR)}=7.3$, or 1 part in 20 million. While it has been possible to get similar contrast ratio performance out of CCDs or standard CMOS devices, using carefully designed observing techniques and instrumentation, the relatively simple approach of direct CID imaging has proved to be encouraging in sub-optimal observing conditions. Some minor improvements in data collection will certainly see contrast ratios in excess of $\Delta{m_v}=19.4, \log{(CR)}=7.8$, or 1 part in 60 million when CID imaging the Sirius field again. Indeed, with a precise photometric calibration and the imaging of other ECR productive fields, a direct contrast ratio greater than $\Delta{m_v}=20, \log{(CR)}=8$ should be achievable with CIDs in the relatively near future. 

\acknowledgments

We are grateful to the anonymous referee for providing suggestions that improved the quality and clarity of this manuscript. This research was supported in part by the American Astronomical Society's Small Research Grant Program, and the Dr. James M. and Sara M. Ortega Astronomy Endowment. DB is grateful to Ben Davies, Dean Hines, Glenn Schneider, and Bill Sparks for useful conversations. 

\bibliographystyle{apj}

\bibliography{../../../batcheldor}

\begin{thebibliography}{53}
\expandafter\ifx\csname natexlab\endcsname\relax\def\natexlab#1{#1}\fi

\bibitem[{{Batcheldor}(2010)}]{2010ApJ...711L.108B}
{Batcheldor}, D. 2010, \apjl, 711, L108

\bibitem[{{Bhaskaran} {et~al.}(2008){Bhaskaran}, {Chapman}, {Pilon}, \&
  {VanGorden}}]{2008SPIE.7055E..19B}
{Bhaskaran}, S., {Chapman}, T., {Pilon}, M., \& {VanGorden}, S. 2008, in
  Society of Photo-Optical Instrumentation Engineers (SPIE) Conference Series,
  Vol. 7055, Society of Photo-Optical Instrumentation Engineers (SPIE)
  Conference Series

\bibitem[{{Blank} {et~al.}(2011){Blank}, {Anglin}, {Beletic}, {Baia}, {Buck},
  {Bhargava}, {Chen}, {Cooper}, {Eads}, {Farris}, {Hall}, {Hodapp}, {Lavelle},
  {Loose}, {Luppino}, {Piquette}, {Ricardo}, {Sprafke}, {Starr}, {Xu}, \&
  {Zandian}}]{2011ASPC..437..383B}
{Blank}, R., {Anglin}, S., {Beletic}, J.~W., {Baia}, Y., {Buck}, S.,
  {Bhargava}, S., {Chen}, J., {Cooper}, D., {Eads}, M., {Farris}, M., {Hall},
  D.~N.~B., {Hodapp}, K.~W., {Lavelle}, W., {Loose}, M., {Luppino}, G.,
  {Piquette}, E., {Ricardo}, R., {Sprafke}, T., {Starr}, B., {Xu}, M., \&
  {Zandian}, M. 2011, in Astronomical Society of the Pacific Conference Series,
  Vol. 437, Solar Polarization 6, ed. J.~R. {Kuhn}, D.~M. {Harrington},
  H.~{Lin}, S.~V. {Berdyugina}, J.~{Trujillo-Bueno}, S.~L. {Keil}, \&
  T.~{Rimmele}, 383

\bibitem[{{Bonnet-Bidaud} {et~al.}(2000){Bonnet-Bidaud}, {Colas}, \&
  {Lecacheux}}]{2000A&amp;A...360..991B}
{Bonnet-Bidaud}, J.~M., {Colas}, F., \& {Lecacheux}, J. 2000, \aap, 360, 991

\bibitem[{{Bonnet-Bidaud} \& {Gry}(1991)}]{1991A&amp;A...252..193B}
{Bonnet-Bidaud}, J.~M. \& {Gry}, C. 1991, \aap, 252, 193

\bibitem[{{Cinzano} {et~al.}(2001){Cinzano}, {Falchi}, \&
  {Elvidge}}]{2001MNRAS.328..689C}
{Cinzano}, P., {Falchi}, F., \& {Elvidge}, C.~D. 2001, \mnras, 328, 689

\bibitem[{{Dunlop} {et~al.}(2003){Dunlop}, {McLure}, {Kukula}, {Baum}, {O'Dea},
  \& {Hughes}}]{2003MNRAS.340.1095D}
{Dunlop}, J.~S., {McLure}, R.~J., {Kukula}, M.~J., {Baum}, S.~A., {O'Dea},
  C.~P., \& {Hughes}, D.~H. 2003, \mnras, 340, 1095

\bibitem[{{Eid}(1995)}]{1995SPIE.2415..292E}
{Eid}, S.~I. 1995, in Society of Photo-Optical Instrumentation Engineers (SPIE)
  Conference Series, Vol. 2415, Charge-Coupled Devices and Solid State Optical
  Sensors V, ed. M.~M. {Blouke}, 292--302

\bibitem[{{Ferrarese} \& {Merritt}(2000)}]{2000ApJ...539L...9F}
{Ferrarese}, L. \& {Merritt}, D. 2000, \apjl, 539, L9

\bibitem[{{Floyd} {et~al.}(2004){Floyd}, {Kukula}, {Dunlop}, {McLure},
  {Miller}, {Percival}, {Baum}, \& {O'Dea}}]{2004MNRAS.355..196F}
{Floyd}, D.~J.~E., {Kukula}, M.~J., {Dunlop}, J.~S., {McLure}, R.~J., {Miller},
  L., {Percival}, W.~J., {Baum}, S.~A., \& {O'Dea}, C.~P. 2004, \mnras, 355,
  196

\bibitem[{{Gebhardt} {et~al.}(2000){Gebhardt}, {Bender}, {Bower}, {Dressler},
  {Faber}, {Filippenko}, {Green}, {Grillmair}, {Ho}, {Kormendy}, {Lauer},
  {Magorrian}, {Pinkney}, {Richstone}, \& {Tremaine}}]{2000ApJ...539L..13G}
{Gebhardt}, K., {Bender}, R., {Bower}, G., {Dressler}, A., {Faber}, S.~M.,
  {Filippenko}, A.~V., {Green}, R., {Grillmair}, C., {Ho}, L.~C., {Kormendy},
  J., {Lauer}, T.~R., {Magorrian}, J., {Pinkney}, J., {Richstone}, D., \&
  {Tremaine}, S. 2000, \apjl, 539, L13

\bibitem[{{Graham} {et~al.}(2001){Graham}, {Erwin}, {Caon}, \&
  {Trujillo}}]{2001ApJ...563L..11G}
{Graham}, A.~W., {Erwin}, P., {Caon}, N., \& {Trujillo}, I. 2001, \apjl, 563,
  L11

\bibitem[{{Ingraham} {et~al.}(2012){Ingraham}, {Doyon}, {Lafreni{\`e}re}, \&
  {Beaulieu}}]{2012PASP..124..454I}
{Ingraham}, P., {Doyon}, R., {Lafreni{\`e}re}, D., \& {Beaulieu}, M. 2012,
  \pasp, 124, 454

\bibitem[{{Jahnke} \& {Macci{\`o}}(2011)}]{2011ApJ...734...92J}
{Jahnke}, K. \& {Macci{\`o}}, A.~V. 2011, \apj, 734, 92

\bibitem[{{Kauffmann} {et~al.}(2003){Kauffmann}, {Heckman}, {Tremonti},
  {Brinchmann}, {Charlot}, {White}, {Ridgway}, {Brinkmann}, {Fukugita}, {Hall},
  {Ivezi{\'c}}, {Richards}, \& {Schneider}}]{2003MNRAS.346.1055K}
{Kauffmann}, G., {Heckman}, T.~M., {Tremonti}, C., {Brinchmann}, J., {Charlot},
  S., {White}, S.~D.~M., {Ridgway}, S.~E., {Brinkmann}, J., {Fukugita}, M.,
  {Hall}, P.~B., {Ivezi{\'c}}, {\v Z}., {Richards}, G.~T., \& {Schneider},
  D.~P. 2003, \mnras, 346, 1055

\bibitem[{{Kimble} {et~al.}(1995){Kimble}, {Chen}, {Haas}, {Norton}, {Payne},
  {Carbone}, \& {Corba}}]{1995SPIE.2518..397K}
{Kimble}, R.~A., {Chen}, P.~C., {Haas}, J.~P., {Norton}, T.~J., {Payne}, L.~J.,
  {Carbone}, J., \& {Corba}, M. 1995, in Society of Photo-Optical
  Instrumentation Engineers (SPIE) Conference Series, Vol. 2518, EUV, X-Ray,
  and Gamma-Ray Instrumentation for Astronomy VI, ed. O.~H. {Siegmund} \& J.~V.
  {Vallerga}, 397--409

\bibitem[{{Kimble} {et~al.}(1999){Kimble}, {Norton}, \&
  {Morrissey}}]{1999ASPC..164..392K}
{Kimble}, R.~A., {Norton}, T.~J., \& {Morrissey}, P.~F. 1999, in Astronomical
  Society of the Pacific Conference Series, Vol. 164, Ultraviolet-Optical Space
  Astronomy Beyond HST, ed. J.~A. {Morse}, J.~M. {Shull}, \& A.~L. {Kinney},
  392

\bibitem[{{Kleinfelder}(2004)}]{2004SPIE.5274..194K}
{Kleinfelder}, S. 2004, in Society of Photo-Optical Instrumentation Engineers
  (SPIE) Conference Series, Vol. 5274, Microelectronics: Design, Technology,
  and Packaging, ed. D.~{Abbott}, K.~{Eshraghian}, C.~A. {Musca},
  D.~{Pavlidis}, \& N.~{Weste}, 194--205

\bibitem[{{Kormendy} \& {Richstone}(1995)}]{1995ARA&A..33..581K}
{Kormendy}, J. \& {Richstone}, D. 1995, \araa, 33, 581

\bibitem[{{Kroupa}(2001)}]{2001MNRAS.322..231K}
{Kroupa}, P. 2001, \mnras, 322, 231

\bibitem[{{Lafreni{\`e}re} {et~al.}(2007){Lafreni{\`e}re}, {Marois}, {Doyon},
  {Nadeau}, \& {Artigau}}]{2007ApJ...660..770L}
{Lafreni{\`e}re}, D., {Marois}, C., {Doyon}, R., {Nadeau}, D., \& {Artigau},
  {\'E}. 2007, \apj, 660, 770

\bibitem[{{Linfield}(2003)}]{2003SPIE.4852..443L}
{Linfield}, R.~P. 2003, in Society of Photo-Optical Instrumentation Engineers
  (SPIE) Conference Series, Vol. 4852, Interferometry in Space, ed. M.~{Shao},
  443--450

\bibitem[{{Liu} {et~al.}(2005){Liu}, {Zhou}, {Ma}, {Wu}, {Yang}, {Li}, \&
  {Chen}}]{2005AJ....129.2628L}
{Liu}, Y., {Zhou}, X., {Ma}, J., {Wu}, H., {Yang}, Y., {Li}, J., \& {Chen}, J.
  2005, \aj, 129, 2628

\bibitem[{{Lowrance} {et~al.}(2005){Lowrance}, {Becklin}, {Schneider},
  {Kirkpatrick}, {Weinberger}, {Zuckerman}, {Dumas}, {Beuzit}, {Plait},
  {Malumuth}, {Heap}, {Terrile}, \& {Hines}}]{2005AJ....130.1845L}
{Lowrance}, P.~J., {Becklin}, E.~E., {Schneider}, G., {Kirkpatrick}, J.~D.,
  {Weinberger}, A.~J., {Zuckerman}, B., {Dumas}, C., {Beuzit}, J.-L., {Plait},
  P., {Malumuth}, E., {Heap}, S., {Terrile}, R.~J., \& {Hines}, D.~C. 2005,
  \aj, 130, 1845

\bibitem[{{Magorrian} {et~al.}(1998){Magorrian}, {Tremaine}, {Richstone},
  {Bender}, {Bower}, {Dressler}, {Faber}, {Gebhardt}, {Green}, {Grillmair},
  {Kormendy}, \& {Lauer}}]{1998AJ....115.2285M}
{Magorrian}, J., {Tremaine}, S., {Richstone}, D., {Bender}, R., {Bower}, G.,
  {Dressler}, A., {Faber}, S.~M., {Gebhardt}, K., {Green}, R., {Grillmair}, C.,
  {Kormendy}, J., \& {Lauer}, T. 1998, \aj, 115, 2285

\bibitem[{{Marconi} \& {Hunt}(2003)}]{2003ApJ...589L..21M}
{Marconi}, A. \& {Hunt}, L.~K. 2003, \apjl, 589, L21

\bibitem[{{Marois} {et~al.}(2006){Marois}, {Lafreni{\`e}re}, {Doyon},
  {Macintosh}, \& {Nadeau}}]{2006ApJ...641..556M}
{Marois}, C., {Lafreni{\`e}re}, D., {Doyon}, R., {Macintosh}, B., \& {Nadeau},
  D. 2006, \apj, 641, 556

\bibitem[{{Marshall} {et~al.}(2001){Marshall}, {Ohki}, {McInnis}, {Ninkov}, \&
  {Carbone}}]{2001RScI...72..713M}
{Marshall}, F.~J., {Ohki}, T., {McInnis}, D., {Ninkov}, Z., \& {Carbone}, J.
  2001, Review of Scientific Instruments, 72, 713

\bibitem[{{Martel} {et~al.}(2003){Martel}, {Ford}, {Tran}, {Illingworth},
  {Krist}, {White}, {Sparks}, {Gronwall}, {Cross}, {Hartig}, {Clampin},
  {Ardila}, {Bartko}, {Ben{\'{\i}}tez}, {Blakeslee}, {Bouwens}, {Broadhurst},
  {Brown}, {Burrows}, {Cheng}, {Feldman}, {Franx}, {Golimowski}, {Infante},
  {Kimble}, {Lesser}, {McCann}, {Menanteau}, {Meurer}, {Miley}, {Postman},
  {Rosati}, {Sirianni}, {Tsvetanov}, \& {Zheng}}]{2003AJ....125.2964M}
{Martel}, A.~R., {Ford}, H.~C., {Tran}, H.~D., {Illingworth}, G.~D., {Krist},
  J.~E., {White}, R.~L., {Sparks}, W.~B., {Gronwall}, C., {Cross}, N.~J.~G.,
  {Hartig}, G.~F., {Clampin}, M., {Ardila}, D.~R., {Bartko}, F.,
  {Ben{\'{\i}}tez}, N., {Blakeslee}, J.~P., {Bouwens}, R.~J., {Broadhurst},
  T.~J., {Brown}, R.~A., {Burrows}, C.~J., {Cheng}, E.~S., {Feldman}, P.~D.,
  {Franx}, M., {Golimowski}, D.~A., {Infante}, L., {Kimble}, R.~A., {Lesser},
  M.~P., {McCann}, W.~J., {Menanteau}, F., {Meurer}, G.~R., {Miley}, G.~K.,
  {Postman}, M., {Rosati}, P., {Sirianni}, M., {Tsvetanov}, Z.~I., \& {Zheng},
  W. 2003, \aj, 125, 2964

\bibitem[{{McCreight} \& {Goebel}(1981)}]{1981ApOpt..20.3189M}
{McCreight}, C.~R. \& {Goebel}, J.~H. 1981, \ao, 20, 3189

\bibitem[{{McCreight} {et~al.}(1986){McCreight}, {McKelvey}, {Goebel},
  {Anderson}, \& {Lee}}]{1986SPIE..686...66M}
{McCreight}, C.~R., {McKelvey}, M.~E., {Goebel}, J.~H., {Anderson}, G.~M., \&
  {Lee}, J.~H. 1986, in Society of Photo-Optical Instrumentation Engineers
  (SPIE) Conference Series, Vol. 686, Infrared detectors, sensors, and focal
  plane arrays, 66--75

\bibitem[{Michon(1974)}]{g1974method}
Michon, G. 1974, Method and apparatus for sensing radiation and providing
  electrical readout, uS Patent 3,786,263

\bibitem[{Michon(1987)}]{michon1987cid}
---. 1987, CID image sensor with a preamplifier for each sensing array row, uS
  Patent 4,689,688

\bibitem[{{Miller} \& {Doughty}(2004)}]{2004SPIE.5167...94M}
{Miller}, K.~B. \& {Doughty}, K.~L. 2004, in Society of Photo-Optical
  Instrumentation Engineers (SPIE) Conference Series, Vol. 5167, Focal Plane
  Arrays for Space Telescopes, ed. T.~J. {Grycewicz} \& C.~R. {McCreight},
  94--100

\bibitem[{{Morrissey} {et~al.}(1998){Morrissey}, {Norton}, \&
  {Kimble}}]{1998SPIE.3356.1036M}
{Morrissey}, P.~F., {Norton}, T.~J., \& {Kimble}, R.~A. 1998, in Society of
  Photo-Optical Instrumentation Engineers (SPIE) Conference Series, Vol. 3356,
  Space Telescopes and Instruments V, ed. P.~Y. {Bely} \& J.~B. {Breckinridge},
  1036--1045

\bibitem[{{Ninkov} {et~al.}(1994{\natexlab{a}}){Ninkov}, {Tang}, {Backer},
  {Easton}, \& {Carbone}}]{1994SPIE.2198..868N}
{Ninkov}, Z., {Tang}, C., {Backer}, B., {Easton}, R.~L., \& {Carbone}, J.
  1994{\natexlab{a}}, in Society of Photo-Optical Instrumentation Engineers
  (SPIE) Conference Series, Vol. 2198, Instrumentation in Astronomy VIII, ed.
  D.~L. {Crawford} \& E.~R. {Craine}, 868--876

\bibitem[{{Ninkov} {et~al.}(1994{\natexlab{b}}){Ninkov}, {Tang}, \&
  {Easton}}]{1994SPIE.2172..180N}
{Ninkov}, Z., {Tang}, C., \& {Easton}, Jr., R.~L. 1994{\natexlab{b}}, in
  Society of Photo-Optical Instrumentation Engineers (SPIE) Conference Series,
  Vol. 2172, Charge-Coupled Devices and Solid State Optical Sensors IV, ed.
  M.~M. {Blouke}, 180--186

\bibitem[{{Oudmaijer} \& {Parr}(2010)}]{2010MNRAS.405.2439O}
{Oudmaijer}, R.~D. \& {Parr}, A.~M. 2010, \mnras, 405, 2439

\bibitem[{{Reggiani} \& {Meyer}(2013)}]{2013A&amp;A...553A.124R}
{Reggiani}, M. \& {Meyer}, M.~R. 2013, \aap, 553, A124

\bibitem[{{Schneider} {et~al.}(2001){Schneider}, {Becklin}, {Smith},
  {Weinberger}, {Silverstone}, \& {Hines}}]{2001AJ....121..525S}
{Schneider}, G., {Becklin}, E.~E., {Smith}, B.~A., {Weinberger}, A.~J.,
  {Silverstone}, M., \& {Hines}, D.~C. 2001, \aj, 121, 525

\bibitem[{{Schneider} {et~al.}(2006){Schneider}, {Silverstone}, {Hines},
  {Augereau}, {Pinte}, {M{\'e}nard}, {Krist}, {Clampin}, {Grady}, {Golimowski},
  {Ardila}, {Henning}, {Wolf}, \& {Rodmann}}]{2006ApJ...650..414S}
{Schneider}, G., {Silverstone}, M.~D., {Hines}, D.~C., {Augereau}, J.-C.,
  {Pinte}, C., {M{\'e}nard}, F., {Krist}, J., {Clampin}, M., {Grady}, C.,
  {Golimowski}, D., {Ardila}, D., {Henning}, T., {Wolf}, S., \& {Rodmann}, J.
  2006, \apj, 650, 414

\bibitem[{{Schneider} {et~al.}(2010){Schneider}, {Silverstone}, {Stobie},
  {Rhee}, \& {Hines}}]{2010hstc.workE..15S}
{Schneider}, G., {Silverstone}, M.~D., {Stobie}, E., {Rhee}, J.~H., \& {Hines},
  D.~C. 2010, in Hubble after SM4. Preparing JWST

\bibitem[{{Schramm} {et~al.}(2008){Schramm}, {Wisotzki}, \&
  {Jahnke}}]{2008A&amp;A...478..311S}
{Schramm}, M., {Wisotzki}, L., \& {Jahnke}, K. 2008, \aap, 478, 311

\bibitem[{{Silverman} {et~al.}(2008){Silverman}, {Mainieri}, {Lehmer},
  {Alexander}, {Bauer}, {Bergeron}, {Brandt}, {Gilli}, {Hasinger}, {Schneider},
  {Tozzi}, {Vignali}, {Koekemoer}, {Miyaji}, {Popesso}, {Rosati}, \&
  {Szokoly}}]{2008ApJ...675.1025S}
{Silverman}, J.~D., {Mainieri}, V., {Lehmer}, B.~D., {Alexander}, D.~M.,
  {Bauer}, F.~E., {Bergeron}, J., {Brandt}, W.~N., {Gilli}, R., {Hasinger}, G.,
  {Schneider}, D.~P., {Tozzi}, P., {Vignali}, C., {Koekemoer}, A.~M., {Miyaji},
  T., {Popesso}, P., {Rosati}, P., \& {Szokoly}, G. 2008, \apj, 675, 1025

\bibitem[{{Sirianni} {et~al.}(2002){Sirianni}, {Nota}, {De Marchi},
  {Leitherer}, \& {Clampin}}]{2002ApJ...579..275S}
{Sirianni}, M., {Nota}, A., {De Marchi}, G., {Leitherer}, C., \& {Clampin}, M.
  2002, \apj, 579, 275

\bibitem[{{Soummer} {et~al.}(2015){Soummer}, {Choquet}, {Pueyo}, {Brendan
  Hagan}, {Gofas-Salas}, {Rajan}, {Perrin}, {Chen}, {Debes}, {Golimowski},
  {Hines}, {Schneider}, {N'Diaye}, {Mawet}, {Marois}, \&
  {Barman}}]{2015AAS...22534921S}
{Soummer}, R., {Choquet}, E., {Pueyo}, L., {Brendan Hagan}, J., {Gofas-Salas},
  E., {Rajan}, A., {Perrin}, M.~D., {Chen}, C., {Debes}, J.~H., {Golimowski},
  D.~A., {Hines}, D.~C., {Schneider}, G., {N'Diaye}, M., {Mawet}, D., {Marois},
  C., \& {Barman}, T. 2015, in American Astronomical Society Meeting Abstracts,
  Vol. 225, American Astronomical Society Meeting Abstracts, 349.21

\bibitem[{{Sparks} \& {Ford}(2002)}]{2002ApJ...578..543S}
{Sparks}, W.~B. \& {Ford}, H.~C. 2002, \apj, 578, 543

\bibitem[{{Stark} \& {Kuchner}(2008)}]{2008ApJ...686..637S}
{Stark}, C.~C. \& {Kuchner}, M.~J. 2008, \apj, 686, 637

\bibitem[{{Tuomi} {et~al.}(2013){Tuomi}, {Jones}, {Jenkins}, {Tinney},
  {Butler}, {Vogt}, {Barnes}, {Wittenmyer}, {O'Toole}, {Horner}, {Bailey},
  {Carter}, {Wright}, {Salter}, \& {Pinfield}}]{2013A&amp;A...551A..79T}
{Tuomi}, M., {Jones}, H.~R.~A., {Jenkins}, J.~S., {Tinney}, C.~G., {Butler},
  R.~P., {Vogt}, S.~S., {Barnes}, J.~R., {Wittenmyer}, R.~A., {O'Toole}, S.,
  {Horner}, J., {Bailey}, J., {Carter}, B.~D., {Wright}, D.~J., {Salter},
  G.~S., \& {Pinfield}, D. 2013, \aap, 551, A79

\bibitem[{{Weinberger} {et~al.}(1999){Weinberger}, {Becklin}, {Schneider},
  {Smith}, {Lowrance}, {Silverstone}, {Zuckerman}, \&
  {Terrile}}]{1999ApJ...525L..53W}
{Weinberger}, A.~J., {Becklin}, E.~E., {Schneider}, G., {Smith}, B.~A.,
  {Lowrance}, P.~J., {Silverstone}, M.~D., {Zuckerman}, B., \& {Terrile}, R.~J.
  1999, \apjl, 525, L53

\bibitem[{Welsch {et~al.}(2006)Welsch, Bravin, Burel, \&
  Lefevre}]{Welsch:2006ef}
Welsch, C.~P., Bravin, E., Burel, B., \& Lefevre, T. 2006, Compact Linear
  Collider Notes, 18

\bibitem[{{Zarnowski} {et~al.}(1996){Zarnowski}, {Carbone}, \&
  {Pace}}]{1996SPIE.2654...29Z}
{Zarnowski}, J.~J., {Carbone}, J., \& {Pace}, M.~A. 1996, in Society of
  Photo-Optical Instrumentation Engineers (SPIE) Conference Series, Vol. 2654,
  Solid State Sensor Arrays and CCD Cameras, ed. C.~N. {Anagnostopoulos}, M.~M.
  {Blouke}, \& M.~P. {Lesser}, 29--37

\bibitem[{Zarnowski {et~al.}(1993)Zarnowski, Eid, Arnold, Pace, Carbone, \&
  Williams}]{doi:10.1117/12.148589}
Zarnowski, J.~J., Eid, S.~I., Arnold, F.~S., Pace, M.~A., Carbone, J., \&
  Williams, B. 1993, Performance of a large-format charge-injection device

\end{thebibliography}

% Tables.

% Figures.

%% This figure uses \includegraphics to scale and rotate the still frame

\end{document}